\documentclass[prd,aps,reprint,twocolumn,preprintnumbers,amsmath,amssymb,superscriptaddress]{revtex4-2}

\usepackage{xcolor}
\usepackage{soul}
\usepackage{graphicx}% Include figure files
\usepackage{dcolumn}% Align table columns on decimal point
\usepackage{bm}% bold math
\usepackage{mathrsfs}
\usepackage{amsmath}
\def\lesssim{\ \raise.3ex\hbox{$<$}\kern-0.8em\lower.7ex\hbox{$\sim$}\ }
\def\gesim{\ \raise.3ex\hbox{$>$}\kern-0.8em\lower.7ex\hbox{$\sim$}\ }

\begin{document}

\preprint{RIKEN-iTHEMS-Report-24}
\title{Non-relativistic trace anomaly and equation of state in dense fermionic matter}

\author{Hiroyuki Tajima}
\affiliation{Department of Physics, Graduate School of Science, The University of Tokyo, Tokyo 113-0033, Japan}

\author{Kei Iida}
\affiliation{Department of Mathematics and Physics, Kochi University, 780-8520, Japan}

\author{Haozhao Liang}
\affiliation{Department of Physics, Graduate School of Science, The University of Tokyo, Tokyo 113-0033, Japan}
\affiliation{RIKEN Interdisciplinary Theoretical and Mathematical Sciences Program, Wako 351-0198, Japan}

\date{\today}

\begin{abstract}
We theoretically investigate a non-relativistic trace anomaly and its impact on the low-temperature equation of state in spatially one-dimensional three-component fermionic systems with a three-body interaction, which exhibit a non-trivial three-body crossover from a bound trimer gas to dense fermionic matter with increasing density.
By applying the $G$-matrix approach to the three-body interaction, we obtain the analytical expression for the ground-state equation of state relevant to the high-density degenerate regime and thereby address how the three-body contact or, equivalently, the trace anomaly emerges. 
The analytical results are compared with the recent quantum Monte Carlo data.
Our study of the trace anomaly and the sound speed could have some relevance to the physics of hadron-quark crossover in compact stars.
\end{abstract}

\maketitle
\par

\section{Introduction}
The recent development of astrophysical observations enables us to address 
the fundamental question of how matter behaves at extremely high density.
Indeed, masses and radii of neutron stars have been simultaneously deduced from 
gravitational waves observed from a binary neutron star merger~\cite{PhysRevLett.119.161101}. 
By incorporating such information and the presence of a heavy neutron star~\cite{demorest2010two} into a Bayesian analysis, the equation of state of neutron star matter has been determined, such that the speed of sound is marginally peaked at several times of the normal nuclear density $\rho_0=0.16$ fm$^{-3}$~\cite{PhysRevLett.121.161101}. 
In a manner that is consistent with this equation of state, nowadays we are in a position to theoretically construct the equation of state of matter at densities significantly higher than $\rho_0$.

In such an extremely dense environment, hadrons, which consist basically of three quarks, overlap with each other and can no longer be regarded as point-like particles.  Eventually, neutron star matter
is expected to be governed by the quark degrees of freedom in the form of color superconducting quark matter~\cite{RevModPhys.80.1455}.
It is nevertheless difficult to figure out how nuclear matter, which is relatively well-known, changes into such quark matter as density increases.
The above-mentioned empirical equation of state of neutron star matter invokes
the so-called hadron-quark crossover scenario~\cite{PhysRevLett.82.3956,baym2018hadrons,kojo2021qcd}, where nuclear matter, if compressed, would undergo a crossover toward quark matter rather than a phase transition~\cite{PhysRevLett.129.181101,PhysRevD.106.103027}.
While its microscopic mechanism is still elusive in the presence of
the sign problem inherent in lattice simulations of finite-density quantum chromodynamics (QCD),
a recent lattice simulation of finite-density two-color QCD~\cite{iida2020two,10.1093/ptep/ptac137}, which is free of the sign problem, indicates a peak of the speed of sound in the density region, where the diquark condensate gradually changes in a similar way to 
the Bose-Einstein condensation (BEC)-to-Bardeen-Cooper-Schrieffer (BCS) crossover~\cite{PhysRev.186.456,nozieres1985bose,leggett2008diatomic,chen2005bcs,zwerger2011bcs,randeria2014crossover,STRINATI20181,ohashi2020bcs} realized in ultracold Fermi atomic gases~\cite{Regal2004PhysRevLett.92.040403,Zwierlein2004PhysRevLett.92.120403,Bartenstein2004PhysRevLett.92.203201}.

In this sense, an alternative promising route to address the microscopic mechanism of the hadron-quark crossover could be via an analog quantum simulation based on ultracold atomic physics~\cite{ketterle2008making,o2011realizing,horikoshi2019cold}.
Thanks to the tunable interactions, adjustable internal degrees of freedom,
and reachable quantum degeneracy through Feshbach resonances, hyperfine states, and state-of-the-art cooling techniques,
respectively,
ultracold atoms offer an ideal platform to investigate quantum many-body physics~\cite{RevModPhys.80.885}.
Indeed, for a non-relativistic one-dimensional (1D) three-component Fermi atom mixture  
with a three-body interaction between different components,
a crossover from a gas of tightly bound trimers to a gas of single atoms with increasing density 
has been pointed out~\cite{PhysRevLett.120.243002};
the thermal equation of state and the minimum of the compressibility (corresponding to the sound velocity peak) in the crossover regime have been reported from a quantum Monte Carlo (QMC) simulation based on the worldline formulation, which is free of
the sign problem~\cite{PhysRevA.102.023313}. 
Such a system can be regarded as a good testing ground for many-body 
theories involving three-body forces, which play a crucial role in low-energy nuclear physics~\cite{RevModPhys.39.745,RevModPhys.85.197}.

Furthermore, 
this 1D system exhibits a trace anomaly due to the broken scale invariance~\cite{PhysRevLett.120.243002,daza2019quantum}.
The same kind of trace anomaly is known to appear in  spatially three-dimensional (3D) dense QCD,
which has been recently discussed in connection with the sound velocity peak~\cite{PhysRevLett.129.252702}.
Note that in both systems, the trace anomaly corresponds to the deviation of the equation of state from the scale-invariant behavior 
in the high density limit.
In the non-relativistic system, which will be studied here, 
the trace anomaly can be expressed in terms of the three-body contact~\cite{PhysRevLett.120.243002,daza2019quantum}, which is a three-body generalization of Tan's two-body contact~\cite{TAN20082952,TAN20082971,TAN20082987} and
characterizes the probability of finding three particles close to each other~\cite{PhysRevLett.106.153005,PhysRevA.86.053633}. 
The numerical values of the three-body contact in the crossover regime have also been obtained from the above-mentioned QMC simulation~\cite{PhysRevA.102.023313}.

It is useful to consider various spatial dimensions and multi-body interactions in the non-relativistic Fermi system of interest here.
The trace anomaly has been experimentally measured in spatially two-dimensional (2D) two-component Fermi  atomic gases with two-body interaction~\cite{PhysRevLett.121.120401,murthy2019quantum}, while it has been theoretically shown that in the high temperature limit,
there is an exact mapping of the two-body anomalous interaction in the 2D model onto the three-body anomalous interaction in the 1D model~\cite{PhysRevA.100.063604}.
Incidentally, both models are asymptotically free: The interactions become asymptotically weaker with increasing density.
Intuitively, this can be understood from the comparison of two energy scales, namely, the Fermi energy $E_{\rm F}$ and the multi-body binding energy $E_{\rm b}$; 
the weak coupling limit is realized when $E_{\rm F}/E_{\rm b}\rightarrow\infty$. 
This property is in contrast to the conventional 3D model where the energy ratio
is characterized by the Fermi momentum $k_{\rm F}$ and the two-body scattering length $a$ in such a way that the high density limit corresponds to the unitarity (i.e., $k_{\rm F}|a|\rightarrow \infty$)~\cite{ohashi2020bcs}.

In considering the crossover mechanism of non-relativistic three-component Fermi mixtures, it is interesting to focus on the analogy with the BEC-BCS crossover in two-component Fermi atomic gases, where tightly bound diatomic molecules are changed into loosely bound Cooper pairs.  Indeed, in three-component Fermi atomic gases, one can expect a similar crossover where tightly bound triatomic molecules are changed into loosely bound trimers called Cooper triples~\cite{PhysRevA.104.053328,PhysRevResearch.4.L012021}.
Such 
a triple state that persists even in the presence of a Fermi sea is a natural extension of the Cooper pairing state and partially consistent with a phenomenological picture of quarkyonic matter that is favorable 
for explaining the sound velocity peak in neutron star matter~\cite{PhysRevLett.122.122701}.
The crossover from baryons to color-singlet Cooper triples has also been discussed theoretically in a semi-relativistic quark model with a phenomenological three-body attraction 
that is responsible for color confinement~\cite{tajima2023density}. 
However, it remains to be investigated 
how the ground-state equation of state is associated with the three-body correlations and the trace anomaly even in the high-density regime of neutron star matter.

In this work,
we theoretically investigate the ground-state equation of state for non-relativistic 1D three-component fermionic matter, which is connected to the trace anomaly just like dense QCD, by using the Brueckner $G$-matrix approach, which is known to successfully describe the ground-state equation of state for asymptotically free 2D Fermi atomic gases~\cite{PhysRevA.84.033607,klawunn2016equation,PhysRevA.107.053313}. 
Remarkably, this approach gives an analytical expression for the equation of state, which in turn well reproduces the equation of state obtained by a QMC simulation~\cite{PhysRevLett.106.110403} and experiments~\cite{PhysRevLett.112.045301,PhysRevLett.114.110403} throughout the 2D BCS-BEC crossover. 
Moreover, the $G$-matrix result for the ground-state energy of a Fermi polaron, namely, an impurity quasiparticle immersed in a Fermi sea, shows an excellent agreement with the exact result in 1D~\cite{10.1063/1.1704798} and experimental results in 2D~\cite{koschorreck2012attractive}. 
We focus on the low-temperature and high-density regime where the QMC simulation is numerically demanding even in this non-relativistic 1D system regardless of the fact that the high-density regime corresponds to the weakly-coupled regime due to the asymptotic freedom.
In particular, we derive an analytical expression for the equation of state and the three-body contact in this system and
elucidate the impact of the trace anomaly on the ground-state equation of state in the presence of strong three-body correlations.

This paper is organized as follows.
In Sec.~\ref{sec:2},
we present a formalism for describing non-relativistic 1D three-component fermions involving 
three-body attractive interaction.
In Sec.~\ref{sec:3}, we discuss the three-body contact and the ground-state equation of state in the high-density regime ($\mu>0$).
We summarize this paper in Sec.~\ref{sec:4}.
Throughout the paper, we take $\hbar=k_{\rm B}=1$, and the system size is set to be unity.

\section{Formalism}
\label{sec:2}
We consider non-relativistic 1D three-component fermions
by starting from the following Hamiltonian~\cite{PhysRevLett.120.243002},
\begin{align}
H=&~
\int\frac{dp}{2\pi}
\sum_{a={\rm r,g,b}}
\epsilon_{p}\psi_{p,a}^\dag \psi_{p,a}\cr
&+g_3
\int\frac{dKdkdqdk'dq'}{(2\pi)^5}
F^\dag(k,q,K) F(k',q',K),
\end{align}
where $\epsilon_{p}=p^2/(2m)$ is the kinetic energy of a fermion with mass $m$, and $\psi_{p,a}$ is the fermion operator with color index $a={\rm r,g,b}$. 
$g_3$ is the contact-type coupling of the three-body force.
The three-body interaction can be expressed in terms of the three-fermion operator, 
\begin{align}
    F(k,q,K)=&~
    \frac{1}{6}\sum_{a_1,a_2,a_3}
    \varepsilon_{a_1a_2a_3}\cr
    &\times
    \psi_{\frac{K}{3}-q,a_1}
\psi_{\frac{K}{3}-k+\frac{q}{2},a_2}
\psi_{\frac{K}{3}+k+\frac{q}{2},a_3},
\end{align}
where $\varepsilon_{a_1a_2a_3}$ is the completely antisymmetric tensor.

\begin{figure}[t]
    \centering
    \includegraphics[width=7cm]{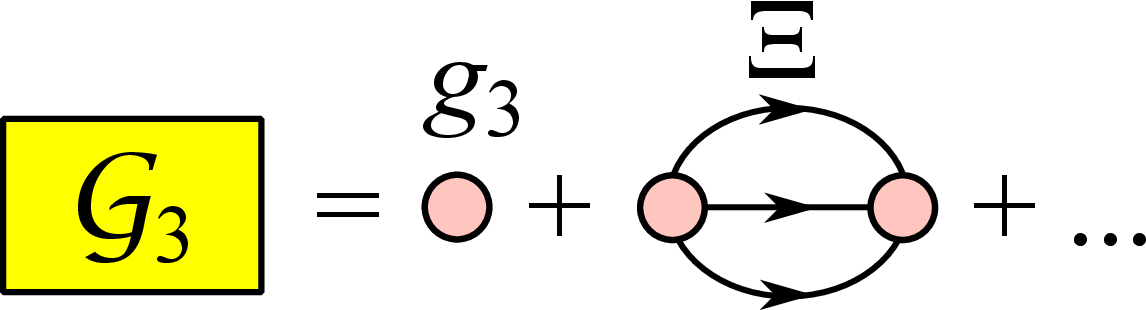}
    \caption{Feynman diagrams for the three-body $G$-matrix $\mathcal{G}_3(K,\omega)$. The circle represents the three-body coupling $g_3$. Three solid lines correspond to the three-body propagator $\Xi(K,\omega)$.}
    \label{fig:1}
\end{figure}

The three-body coupling constant $g_3$ induces a three-body bound state even for infinitesimally small $g_3$ in 1D.
Because of non-perturbative properties of the three-body coupling, we need to sum up an infinite series of the three-body ladder diagrams shown in Fig.~\ref{fig:1} even in the weak-coupling (or high-density) regime.
The three-body $G$-matrix is given by
\begin{align}
\label{eq:gmat}
\mathcal{G}_3(K,\omega)&=\left[\frac{1}{g_3}-\Xi(K,\omega)\right]^{-1},
\end{align}
where
\begin{align}
\label{eq:xi}
    \Xi(K,\omega)=\int \frac{dkdq}{(2\pi)^2}
    \frac{Q(k,q,K)}{\omega-\mathcal{E}_{k,q,K}}
\end{align}
is the three-body propagator with the Pauli-blocking factor $Q(k,q,K)$, and
$\mathcal{E}_{k,q,K}$ is
the kinetic energy of three particles given by 
\begin{align}
\mathcal{E}_{{k},{q},{K}}=\frac{\left(\frac{q}{2}+\frac{K}{3}+k\right)^2}{2m}+\frac{\left(\frac{K}{3}-{q}\right)^2}{2m}
+\frac{\left(\frac{q}{2}+\frac{K}{3}-k\right)^2}{2m}.
\end{align}
The three-body coupling can be characterized by the three-body binding energy $E_{\rm b}$ in vacuum obtained from the pole of the three-body $T$-matrix~\cite{PhysRevResearch.4.L012021}.
Namely, we take $Q(k,q,K)=1$ in Eq.~(\ref{eq:xi}) and obtain
\begin{align}
\label{eq:g3}
    \frac{1}{g_3}&=-\int\frac{dkdq}{(2\pi)^2}\frac{1}{E_{\rm b}+\mathcal{E}_{{k},{q},0}}\cr
    &=-\frac{m}{2\sqrt{3}\pi}
    \ln\left(\frac{mE_{\rm b}+\Lambda^2}{mE_{\rm b}}\right),
\end{align}
where $\Lambda$ is the momentum cutoff.
In this regard, one can find that a usual Hartree-like lowest-order interaction energy $g_3\rho_{\rm r}\rho_{\rm b}\rho_{\rm g}$, where $\rho_{a}$ is the number density of color $a$, vanishes in the limit of $\Lambda\rightarrow \infty$,
indicating that an appropriate regularization of $\Lambda$ is needed even in the high-density limit.

Hereafter, we focus on the color-symmetric case with $\rho_{\rm r}=\rho_{\rm b}=\rho_{\rm g}\equiv \rho/3$, where $\rho$ is the total number density.
Following the idea of the Brueckner Hartree-Fock theory in the presence of a bound state~\cite{PhysRevA.84.033607,klawunn2016equation,PhysRevA.107.053313},
we evaluate the internal energy $E$ as the Hartree-Fock-like expectation value $E=\langle H\rangle$ by replacing $g_3$ with the in-medium effective interaction $\mathcal{G}_3(K=0,\omega=-E_{\rm b})$, that is,
\begin{align}
    E= \frac{1}{3}\rho E_{\rm F}+\frac{1}{27}\mathcal{G}_3(0,-E_{\rm b})\rho^3.
\end{align}
This approximation leads to an analytical expression for the equation of state, which works unexpectedly
well in two-dimensional two-component Fermi atomic gases with attractive interaction throughout the BCS-BEC crossover~\cite{klawunn2016equation,PhysRevA.107.053313}.

We evaluate $\Xi(0,-E_{\rm b})$ in $\mathcal{G}_3(0,-E_{\rm b})$ within the Tamm-Dancoff approximation where low-energy excitations 
below the Fermi energy $E_{\rm F}$ are suppressed at $T=0$~\cite{ring2004nuclear}.
This treatment is similar to the generalized Cooper problem for three-body states~\cite{PhysRevA.86.013628,PhysRevA.96.053614,PhysRevA.104.L041302,PhysRevA.104.053328,PhysRevB.107.024511}.
As in the case of 2D two-component Fermi gases in which 
the states with zero center-of-mass momentum can be regarded as the relevant contribution, 
the states with $K=0$, which correspond to
squeezed Cooper triples, can be relevant in 1D three-component Fermi gases~\cite{PhysRevA.104.053328,PhysRevResearch.4.L012021}.
Because of the internal degrees of freedom associated with constituent fermions in three-body cluster states,
the three-body correlations at $K=0$ involve an ultraviolet divergence with respect to the integration of relative momenta.
Although we do not consider the condensation in the present 1D system, 
the low-momentum correlations play a crucial role at low temperature and high density; indeed, such an approximation shows a good agreement with experiments~\cite{PhysRevLett.112.045301,PhysRevLett.114.110403,koschorreck2012attractive} as well as QMC simulations~\cite{PhysRevLett.100.030401,PhysRevLett.106.153005} as shown in Refs.~\cite{PhysRevA.84.033607,klawunn2016equation,PhysRevA.107.053313}.
On the other hand, our approach cannot reproduce the low-density limit where a gas of tightly bound trimers is realized because we do not consider three-body correlations with $K\neq 0$.
In this regard, we focus on the high-density regime where the contribution with $K=0$ can be expected to be dominant.
By introducing $\bar{q}=\frac{\sqrt{3}}{2}q$ and $r=\sqrt{k^2+\bar{q}^2}$, we obtain 
\begin{align}
    \Xi({0},-E_{\rm b})
    &=-\frac{m}{2\sqrt{3}\pi^2}\int_0^{\Lambda}2\pi r dr\frac{Q\left(k,\frac{2}{\sqrt{3}}\bar{q},0\right)}{mE_{\rm b}+r^2}.
\end{align}
Note here that
$r^2=m\mathcal{E}_{k,q,0}\geq 3mE_{\rm F}$. Then, we find $Q\left(k,\frac{2}{\sqrt{3}}\bar{q},0\right)\equiv Q(r)=\theta\left(r-\sqrt{\frac{3}{2}}k_{\rm F}\right)$ with the Fermi momentum $k_{\rm F}=\frac{\pi\rho}{3}$, leading to
\begin{align}
    \Xi({0},-E_{\rm b})
    &=
    -\frac{m}{2\sqrt{3}\pi}
    \ln\left(\frac{mE_{\rm b}+\Lambda^2}{mE_{\rm b}+\frac{3}{2}k_{\rm F}^2}\right).
\end{align}
Accordingly, we obtain
\begin{align}
\label{eq:gmat2}
    \mathcal{G}_3({0},-E_{\rm b})
    &=-
    \frac{2\sqrt{3}\pi}{m}\left[\ln\left(\frac{mE_{\rm b}+\frac{3}{2}k_{\rm F}^2}{mE_{\rm b}}\right)
    \right]^{-1}.
\end{align}
One can see that $\mathcal{G}_3(0,-E_{\rm b})$ in Eq.~\eqref{eq:gmat2} does not depend on $\Lambda$ in contrast to $g_3$ in Eq.~\eqref{eq:g3}.

Eventually, the internal energy within the present approach reads
\begin{align}
    E
    &=
    \frac{1}{3}\rho E_{\rm F}
    -\frac{4\sqrt{3}}{3\pi}\rho E_{\rm F}
    \frac{1}{\ln\left(1+\frac{3E_{\rm F}}{E_{\rm b}}\right)}.
\end{align}
In particular, 
it is worth mentioning that
in the high-density limit ($E_{\rm F}\gg E_{\rm b}$)
one can obtain
\begin{align}
\label{eq:weak-coupling}
    E\simeq\frac{1}{3}\rho E_{\rm F}
    -\frac{4\sqrt{3}}{3\pi}\rho E_{\rm F}
    /\ln\left(\frac{3E_{\rm F}}{E_{\rm b}}\right),
\end{align}
which is similar to the lowest-order interaction correction in 2D Fermi gases with two-body interaction~\cite{klawunn2016equation,PhysRevA.83.021603}.
While the $G$-matrix approach can be justified in the high-density regime due to the asymptotic freedom,
the logarithmic correction in Eq.~\eqref{eq:weak-coupling} is the consequence of non-perturbative nature of the three-body coupling captured by the infinite ladder resummation in Fig.~\ref{fig:1}.

Some other thermodynamic quantities can be obtained via the thermodynamic identities.
The chemical potential $\mu=\frac{\partial E}{\partial \rho}$ is given by 
\begin{align}
    \frac{\mu}{E_{\rm F}}
     =&~ 1
     -\frac{4\sqrt{3}}{\pi}
     \frac{1}{\ln\left(1+\frac{3E_{\rm F}}{E_{\rm b}}\right)}\cr
     &+
     \frac{8\sqrt{3}}{3\pi }
     \frac{3E_{\rm F}/E_{\rm b}}{\left(1+\frac{3E_{\rm F}}{E_{\rm b}}\right)
     \left[\ln\left(1+\frac{3E_{\rm F}}{E_{\rm b}}\right)\right]^2}.
\end{align}
The pressure $P=\mu\rho-E$ reads
\begin{align}
    P
     =&~ \frac{2}{3}\rho E_{\rm F}
     -\frac{8\sqrt{3}}{3\pi}\rho E_{\rm F}
     \frac{1}{\ln\left(1+\frac{3E_{\rm F}}{E_{\rm b}}\right)}\cr
     &+
     \frac{8\sqrt{3}}{3\pi } \rho E_{\rm F}
     \frac{3E_{\rm F}/E_{\rm b}}{\left(1+\frac{3E_{\rm F}}{E_{\rm b}}\right)
     \left[\ln\left(1+\frac{3E_{\rm F}}{E_{\rm b}}\right)\right]^2}.
\end{align}
In the high-density limit where $E_{\rm b}$ is negligible compared to $E_{\rm F}$,
one can find the scale-invariant result $P=2E$~\cite{PhysRevLett.120.243002}.
However, at lower densities, such a relation is gradually broken due to the trace anomaly, which is equivalent to the three-body contact $C_3=P-2E$~\cite{daza2019quantum}.
We thus obtain
\begin{align}
    C_3
     &=
     \frac{8\sqrt{3}}{3\pi } \rho E_{\rm F}
     \frac{3E_{\rm F}/E_{\rm b}}{\left(1+\frac{3E_{\rm F}}{E_{\rm b}}\right)
     \left[\ln\left(1+\frac{3E_{\rm F}}{E_{\rm b}}\right)\right]^2}.
\end{align}
One can easily find that $C_3$ is positive definite as in the case of conventional Tan's contact~\cite{TAN20082952,TAN20082971,TAN20082987}.

Moreover, one can obtain the squared sound velocity $c_s^2=\frac{1}{m}\left(\frac{\partial P}{\partial \rho}\right)$ as
\begin{align}
    \frac{c_s^2}{v_{\rm F}^2}
     =&~1-\frac{4\sqrt{3}}{\pi}\frac{1}{\ln\left(1+\frac{3E_{\rm F}}{E_{\rm b}}\right)}\cr
     &+\frac{28\sqrt{3}}{\pi}\frac{E_{\rm F}}{E_{\rm b}}
     \frac{1}{\left(1+\frac{3E_{\rm F}}{E_{\rm b}}\right)
     \left[\ln\left(1+\frac{3E_{\rm F}}{E_{\rm b}}\right)\right]^2}\cr
     &-\frac{24\sqrt{3}}{\pi}
     \frac{E_{\rm F}^2}{E_{\rm b}^2}
     \frac{2+\ln\left(1+\frac{3E_{\rm F}}{E_{\rm b}}\right)}{
     \left(1+\frac{3E_{\rm F}}{E_{\rm b}}\right)^2
     \left[\ln\left(1+\frac{3E_{\rm F}}{E_{\rm b}}\right)\right]^3
     },
\end{align}
where $v_{\rm F}=k_{\rm F}/m$ is the Fermi velocity and corresponds to the high-density conformal limit in this system.
We note that at $T=0$, $c_s^2$ is related to the compressibility $\kappa=\frac{1}{\rho}\left(\frac{\partial\rho}{\partial P}\right)$. 
By introducing the non-interacting compressibility $\kappa_0=\rho/v_{\rm F}^2$, we find $\kappa/\kappa_0=v_{\rm F}^2/c_s^2$.

\section{Results}
\label{sec:3}
\begin{figure}[t]
    \centering
    \includegraphics[width=7.5cm]{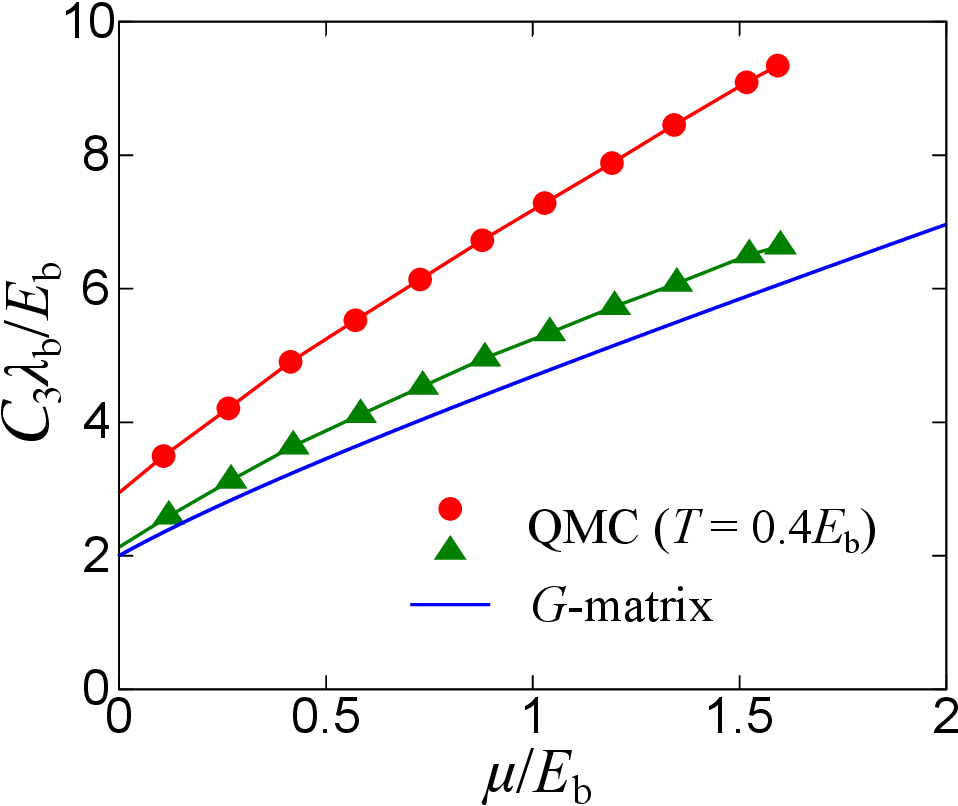}
    \caption{Three-body contact $C_3$ as a function of $\mu/E_{\rm b}$ at $T=0$,
    where $\lambda_{\rm b}=\sqrt{\frac{2\pi}{mE_{\rm b}}}$ is the length scale associated with the three-body binding energy $E_{\rm b}$.
    For comparison, we also show the QMC results at $T=0.4E_{\rm b}$ in Ref.~\cite{PhysRevA.102.023313}, where the circles and triangles are evaluated by two different ways.
    These two results are different due possibly to lattice artifacts.}
    \label{fig:2}
\end{figure}
First, we discuss the three-body contact $C_3$, which arises from the trace anomaly.
In particular, we focus on the regime with $\mu>0$ where the Fermi degeneracy of constituent fermions is important.
Figure~\ref{fig:2} shows $C_3$ as a function of $\mu/E_{\rm b}$
at $T=0$.
One can see that $C_3$ increases with $\mu$, indicating the importance of the low-energy three-body correlations which are reminiscent of squeezed Cooper triples~\cite{PhysRevResearch.4.L012021}. 
We note that this increment is also associated with the increase of $\rho$ as $C_3$ is normalized by the density-independent scales $\lambda_{\rm b}=\sqrt{\frac{2\pi}{mE_{\rm b}}}$ and $E_{\rm b}$ in Fig.~\ref{fig:2}.
If we use the density-dependent scale (e.g., $C_3/\rho E_{\rm F}$), such a quantity vanishes and hence the scale-invariant result $C_3/\rho E_{\rm F}=(P-2E)/\rho E_{\rm F}\rightarrow 0$ is recovered in the high-density limit.  
Even within our simplified approach,
our result is close to the QMC results performed at finite temperature ($T=0.4E_{\rm b}$) in Ref.~\cite{PhysRevA.102.023313}.
In the present $G$-matrix approach where $C_3$ is a little bit underestimated,
we do not consider the three-body correlations with nonzero $K$ and the trimer-trimer interaction~\cite{PhysRevA.99.013615}, which may be the origin of such underestimation of $C_3$.

\begin{figure}[t]
    \centering
    \includegraphics[width=7.5cm]{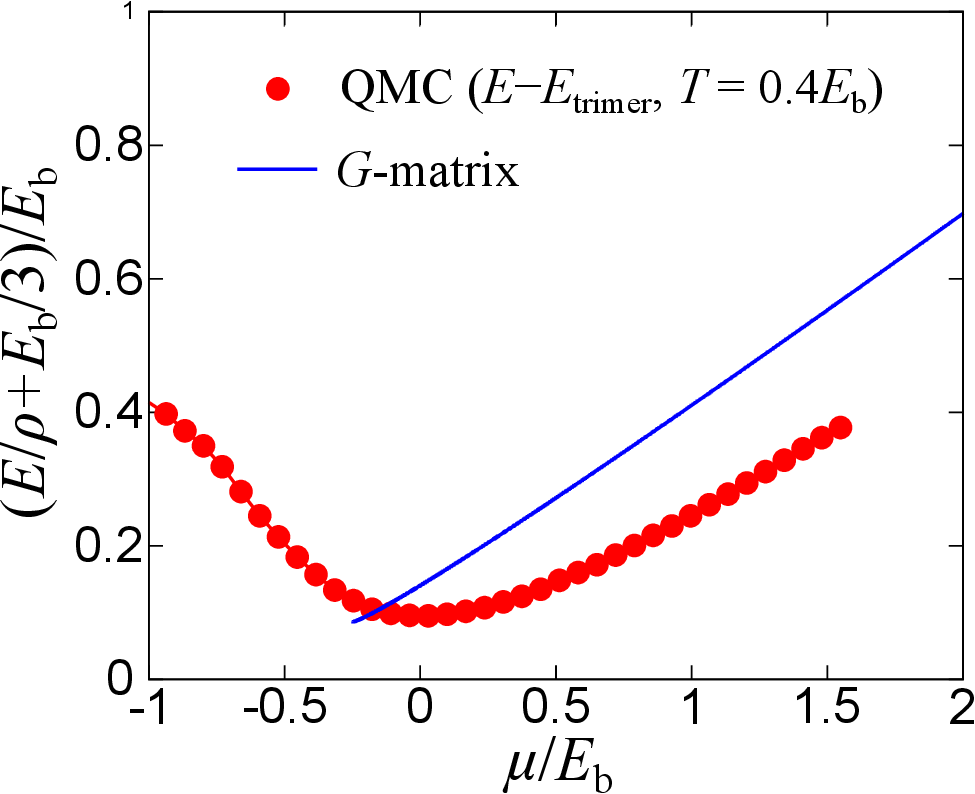}
    \caption{Internal energy $(E/\rho+E_{\rm b}/3)/E_{\rm b}$ as a function $\mu/E_{\rm b}$. For comparison, the QMC result at $T=0.4E_{\rm b}$ in Ref.~\cite{PhysRevA.102.023313} is shown, where the plot has the trimer contribution $E_{\rm trimer}$  subtracted out.}
    \label{fig:3}
\end{figure}

Figure~\ref{fig:3} shows the internal energy per particle $E/\rho+E_{\rm b}/3$. 
In Ref.~\cite{PhysRevA.102.023313}, the QMC results for the internal energy density $E$ that has the the trimer contribution $E_{\rm trimer}=E_{\rm TFG}-\rho E_{\rm b}/3$ subtracted out were reported, where $E_{\rm TFG}$ is the internal energy density of a non-interacting trimer Fermi gas.
Since $E_{\rm TFG}$ is not considered in our calculation,
we compare the $G$-matrix result for $E/\rho+E_{\rm b}/3$ with $(E-E_{\rm trimer})/\rho$ in the QMC simulation.
At $\mu>0$, our result is qualitatively consistent with the QMC result as both results show linear increase with $\mu/E_{\rm b}$. This enhancement is also related to the increase of $\rho$.
The quantity shown in Fig.~\ref{fig:3} indicates the degree to which the system differs from a non-interacting trimer Fermi gas.
In this sense, the result for $C_3$ obtained by the $G$-matrix approach may be regarded as the three-body correlations that cannot be described by the point-like trimer formation.
On the other hand, the trimer-trimer repulsive interaction~\cite{PhysRevA.99.013615} is not considered in our calculation.
This repulsion would act to increase $E$ and thus
lead to further discrepancy between the QMC simulation and the $G$-matrix approach.

\begin{figure}[t]
    \centering
    \includegraphics[width=7.5cm]{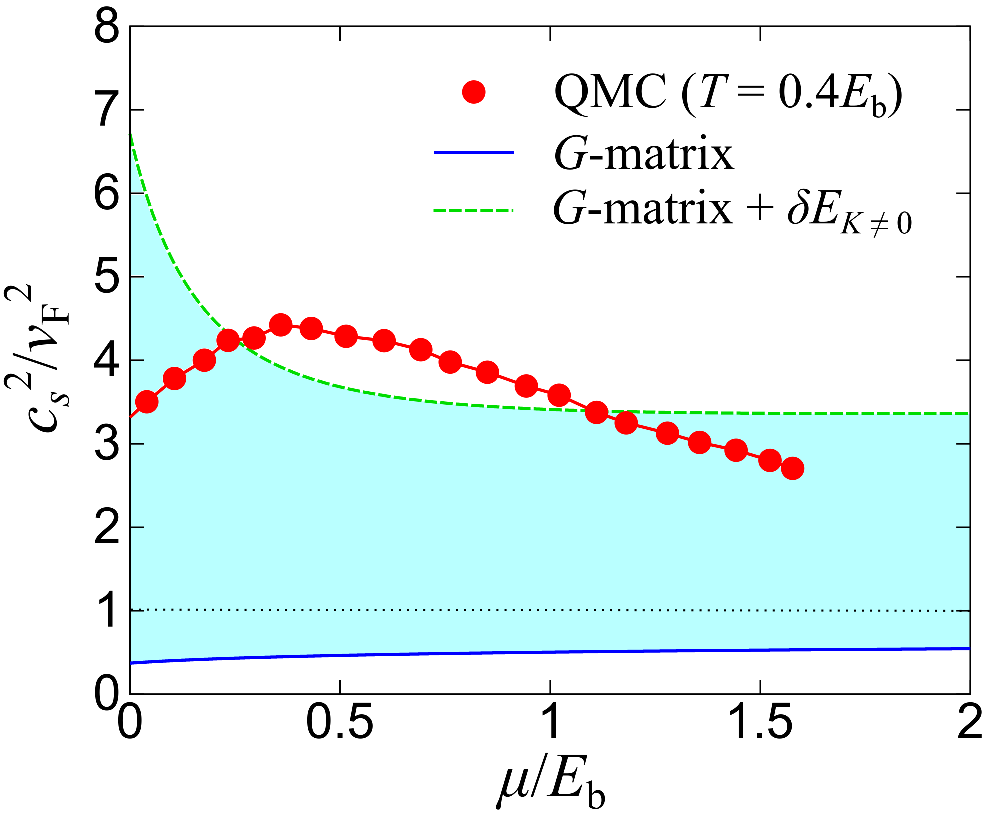}
    \caption{Squared sound velocity $c_s^2/v_{\rm F}^2$ as a function of $\mu/E_{\rm b}$ where $v_{\rm F}$ is the Fermi velocity corresponding to the conformal high-density limit.
    The solid and dashed curves represent the $G$-matrix results with and without the degenerate trimer contribution $\delta c_{K\neq 0}^2$, respectively. The circles show the QMC results obtained from $c_s^2/v_{\rm F}^2=\kappa_0/\kappa$ at $T=0.4E_{\rm b}$. 
    The horizontal dotted line corresponds to the high-density limit ($c_s^2/v_{\rm F}^2=1$).
    }
    \label{fig:4}
\end{figure}

Finally, we examine the squared sound velocity $c_s^2/v_{\rm F}^2$ as shown in Fig.~\ref{fig:4}.
For comparison, we also show the QMC result obtained from the dimensionless compressibility $\kappa/\kappa_0$ at $T=0.4E_{\rm b}$.
Although the QMC result, which is above unity, involves the finite-temperature effect,
the $G$-matrix result for $c_s^2/v_{\rm F}^2$ is well below unity.
To understand this discrepancy, we phenomenologically introduce the three-body correlations with nonzero center-of-mass momenta ($K\neq0$) as $E\rightarrow E+\delta E_{K\neq 0}$ where
\begin{align}
\label{eq:delE}
    \delta E_{K\neq 0}=\int\frac{dK}{2\pi}\frac{K^2}{6m}\theta(K_{\rm T}-K)=\frac{K_{\rm T}^3}{18\pi m}.
\end{align}
In Eq.~\eqref{eq:delE}, $K_{\rm T}=\sqrt{6m(3E_{\rm F}+E_{\rm b})}$ is the effective trimer Fermi momentum.
Then, using the thermodynamic identities,
we find the associated correction to the squared sound velocity $c_s^2\rightarrow c_s^2+\delta c_{s,K\neq 0}^2$, where
\begin{align}
    \delta c_{s,K\neq 0}^2=\frac{k_{\rm F}K_{\rm T}}{2m^2}+\frac{9k_{\rm F}^3}{2m^2K_{\rm T}}.
\end{align}
The dashed curve in Fig.~\ref{fig:4} shows the result for $c_s^2/v_{\rm F}^2$ including the phenomenological three-body correlations 
with $K\neq 0$.  Indeed, it is close to the QMC result.
This indicates the importance of the Pauli pressure of in-medium trimers in addition to the trace anomaly in the high-density regime.
On the other hand, $\delta c_{s,K\neq 0}^2$ does not vanish even in the high-density limit ($E_{\rm F}\gg E_{\rm b}$), whereas $c_{s}^2/v_{\rm F}^2$ should approach unity in the high-density limit.
In this sense, the phenomenological expression for $\delta c_{s,K\neq 0}^2$ 
based on point-like trimer states overestimates the excess of $c_{s}^2$, implying that the non-local Cooper-triple-like correlations with $K\neq 0$ should be taken into account~\cite{PhysRevResearch.4.L012021}.
The trimer-trimer interaction would also play an important role in changing $c_{s}^2$, in addition to $C_3$ and $E$.
Although the $G$-matrix result is known to follow 
$c_s^2/v_{\rm F}^2\simeq 1 -\frac{4\sqrt{3}}{\pi\ln\left(3E_{\rm F}/E_{\rm b}\right)}$, which
is consistent with the equation of state in the BCS-BEC crossover~\cite{klawunn2016equation,PhysRevA.107.053313}, therefore,
a more detailed investigation of the high-density asymptotic behavior of $c_s^2$ is left for an important future work.

\section{Summary}
\label{sec:4}
To summarize, we have investigated the trace anomaly and its impact on the ground-state equation of state for non-relativistic 1D three-component fermions.
By extending the $G$-matrix approach developed for the system with two-body interaction to the system with three-body interaction,
we have obtained the analytical expression for the ground-state equation of state in the presence of three-body correlations that cannot be described by the formation of point-like trimers.
The three-body contact, which results from the non-relativistic trace anomaly, is found to increase with the chemical potential. 
Our results are qualitatively consistent with the recent QMC results at positive chemical potentials even within the simplified approximations adopted here.
We expect that
the Cooper-triple-like three-body correlations appear in the present system. 

As for future perspectives, it is important to consider the three-body correlations with nonzero center-of-mass momenta as well as trimer-trimer interactions for further understanding of the three-body crossover equation of state.
Finite-temperature effects should also be addressed for more quantitative comparison with the QMC calculation.
Moreover, it would be interesting to apply the present approach to hadron-quark crossover by considering a system of constituent quarks interacting via a three-body color-confining force. 

\acknowledgements
The authors thank the members of the Low-Energy Nuclear Theory group at The University of Tokyo for useful discussion on these and related subjects and J. E. Drut for providing information on Ref.~\cite{PhysRevA.102.023313}.
Also, H.T. is grateful to N. Yamamoto for useful discussion at the Workshop ``Thermal Quantum Field Theory" held in KEK.
This research was supported in part by Grants-in-Aid for Scientific
Research provided by JSPS through Nos.~18H05406, 22H01158, 22K13981, and 23H01167. 

\bibliographystyle{apsrev4-2}
\bibliography{reference.bib}

\end{document}